\documentclass[preprint,prd,aps,nofootinbib]{revtex4}
\begin{document}
\begin{center}
\bibliographystyle{article}

{\Large \textsc{On boosted space-times with cosmological constant and
their ultrarelativistic limit}}

\end{center}
\vspace{0.4cm}


\date{\today}

\author{Giampiero Esposito,$^{1,2}$ \thanks{%
Electronic address: giampiero.esposito@na.infn.it} 
Roberto Pettorino,$^{2,1}$
\thanks{%
Electronic address: roberto.pettorino@na.infn.it} and Paolo Scudellaro$^{2,1}$
\thanks{%
Electronic address: scud@na.infn.it}}

\affiliation{${\ }^{1}$Istituto Nazionale di Fisica Nucleare, 
Sezione di Napoli,\\
Complesso Universitario di Monte S. Angelo, Via Cintia, Edificio N', 80126
Napoli, Italy\\
${\ }^{2}$Dipartimento di Scienze Fisiche, Complesso Universitario di Monte
S. Angelo,\\
Via Cintia, Edificio N', 80126 Napoli, Italy}

\begin{abstract}
The problem of deriving a shock-wave geometry with cosmological
constant by boosting a Schwarzschild-de Sitter 
(or anti-de Sitter) black hole is re-examined. Unlike previous work in
the literature, we deal with the {\it exact} Schwarzschild-de Sitter 
(or anti-de Sitter) metric. 
In this exact calculation, where the metric does not depend linearly 
on the mass parameter, we find a singularity of distributional nature
on a null hypersurface, which corresponds to a shock-wave geometry
derived in a fully non-perturbative way. The result agrees with previous
calculations, where the metric had been linearized in the mass parameter.  
\end{abstract}

\maketitle
\bigskip
\vspace{2cm}

\section{Introduction}

The subject of gravitational fields generated by sources which move at
the speed of light has always received much attention in the literature.
In the sixties it was already known that the fields produced by null
sources are plane-fronted gravitational waves
\cite{Pere60, Bonn69}. In Ref. \cite{Aich71},
Aichelburg and Sexl studied the field of a point particle with zero rest
mass moving with the speed of light. They found that the gravitational
field of such a particle is non-vanishing only on a plane containing
the particle and orthogonal to the direction of motion. On this plane
the Riemann tensor has a distributional (Dirac-delta-like) singularity
and is exactly of Petrov type $N$ (i.e. all four principal null directions
of the Weyl spinor, describing the Weyl conformal curvature, 
coincide). For this purpose, the authors of Ref.
\cite{Aich71} used in part a set of Lorentz transformations in the
ultrarelativistic limit.

Since then, other authors considered `boosting'
the Kerr or Kerr--Newman solutions \cite{Ferr90, Bala95, Bala96, Lous92},
while the work by Hotta and Tanaka 
in Ref. \cite{Hott93}, motivated by the analysis
of quantum effects of gravitons in de Sitter space-time, studied the problem
of boosting the Schwarzschild--de Sitter metric to a similar limit
(see also the work in Ref. \cite{Podo97}). For this purpose, the authors
of Refs. \cite{Hott93, Podo97} approximated the
Schwarzschild--de Sitter metric by a first-order perturbation of
de Sitter, i.e.
\begin{equation}
ds^{2} \approx -\left(1-{2m\over r}-{r^{2}\over a^{2}}\right)dt^{2}
+\left(1-{r^{2}\over a^{2}}\right)^{-1}
\left[1+\left(1-{r^{2}\over a^{2}}\right)^{-1}{2m\over r}\right]dr^{2}
+r^{2}(d\theta^{2}+\sin^{2}\theta d\phi^{2}).
\label{(1)}
\end{equation}
They then found that a suitable change of coordinates, combined with
the ultra-relativistic limit (cf. below), lead to a resulting space-time
which differs from de Sitter space-time only by the inclusion of an
impulsive wave located on a null hypersurface.

However, since the Einstein theory is ruled by non-linear effects in the
first place, we remark that an exact analysis relying upon the full
Schwarzschild--de Sitter metric, which depends non-linearly on the
mass parameter, would be desirable if only possible (as will be shown
below). Thus, in our problem, we start from the standard form of 
the metric for a Schwarzschild-de Sitter space-time, i.e.
\begin{equation}
ds^{2}=-\left(1-{2m \over r}-{r^{2}\over a^{2}}\right)dt^{2}
+{dr^{2}\over \left(1-{2m \over r}-{r^{2}\over a^{2}}\right)}
+r^{2}(d\theta^{2}+\sin^{2} \theta d\phi^{2}),
\label{(2)}
\end{equation}
where $a^{2}$ is a parameter 
related to the cosmological constant $\Lambda$ by
$\Lambda={3\over a^{2}}$. Moreover, we know that a de Sitter space-time
in four dimensions can be viewed as a four-dimensional hyperboloid 
embedded in five-dimensional Minkowski space-time. Its metric reads
\begin{equation}
ds_{\rm dS}^{2}=-dZ_{0}^{2}+\sum_{i=1}^{4}dZ_{i}^{2},
\label{(3)}
\end{equation}
with coordinates satisfying the hyperboloid constraint
\begin{equation}
a^{2}=-Z_{0}^{2}+\sum_{i=1}^{4}Z_{i}^{2},
\label{(4)}
\end{equation}
so that the parameter $a$ is just the `radius' of this hyperboloid.

The plan of our paper is as follows.
Section 2 studies the metric (2) in the $Z_{\mu}$ coordinates above.
Section 3 performs a boost in the $Z_{1}$-direction, while section 4
is devoted to a similar analysis in Schwarzschild-anti-de Sitter 
space-time. Section 5 derives in detail a shock-wave geometry from
the ultrarelativistic limit of such boosted space-times with cosmological
constant, while concluding remarks are presented in section 6.

\section{First change of coordinates}

The work in \cite{Hott93} exploits the relation between the 
$Z_{\mu}$ coordinates in Eqs. (3) and (4) and the spherical static coordinates
$(t,r,\theta,\phi)$:
\begin{equation}
Z_{0} \equiv \sqrt{a^{2}-r^{2}} \; \sinh (t/a),
\label{(5)}
\end{equation}
\begin{equation}
Z_{4} \equiv \pm \sqrt{a^{2}-r^{2}} \; \cosh(t/a),
\label{(6)}
\end{equation}
\begin{equation}
Z_{1} \equiv r \cos \theta,
\label{(7)}
\end{equation}
\begin{equation}
Z_{2} \equiv r \sin \theta \cos \phi,
\label{(8)}
\end{equation}
\begin{equation}
Z_{3} \equiv r\sin \theta \sin \phi.
\label{(9)}
\end{equation}
The key point of our analysis is to rewrite the {\it exact} metric (2)
in the $Z_{\mu}$ coordinates, without resorting to a perturbative 
expansion up to terms linear in the black hole mass $m$. Thus, on defining
\begin{equation}
f^{2} \equiv a^{2}-r^{2}=Z_{4}^{2}-Z_{0}^{2},
\label{(10)}
\end{equation}
\begin{equation}
F_{m} \equiv 1-{2a^{2}m \over f^{2}r}
-{a^{2}/r^{2}\over \left(1-{2a^{2}m \over f^{2}r}\right)},
\label{(11)}
\end{equation}
\begin{equation}
Q \equiv 1+{2Z_{0}^{2}\over f^{2}},
\label{(12)}
\end{equation}
we re-express the Schwarzschild-de Sitter metric in the form (see
details in Appendix A)
\begin{equation}
ds^{2}=h_{00}dZ_{0}^{2}+h_{44}dZ_{4}^{2}+2h_{04}dZ_{0}dZ_{4}
+dZ_{1}^{2}+dZ_{2}^{2}+dZ_{3}^{2},
\label{(13)}
\end{equation}
where
\begin{equation}
h_{00} \equiv -{1\over 2}(Q-1)F_{m}
-\left(1-{2a^{2}m \over f^{2}r}\right)-{Z_{0}^{2}\over r^{2}},
\label{(14)}
\end{equation}
\begin{equation}
h_{44} \equiv -{1\over 2}(Q+1)F_{m}
+\left(1-{2a^{2}m \over f^{2}r}\right)-{Z_{4}^{2}\over r^{2}},
\label{(15)}
\end{equation}
\begin{equation}
h_{04} \equiv {Z_{0}Z_{4}\over f^{2}}F_{m}
+{Z_{0}Z_{4}\over r^{2}},
\label{(16)}
\end{equation}
with the ratio ${a^{2}m \over f^{2}r}$ given by
\begin{equation}
{a^{2}m \over f^{2}r}={a^{2}m \over (Z_{4}^{2}-Z_{0}^{2})
\sqrt{a^{2}+Z_{0}^{2}-Z_{4}^{2}}}.
\label{(17)}
\end{equation}

\section{The boost in the $Z_{1}$-direction}

The next step, following \cite{Hott93}, is to set 
\begin{equation}
m=p \sqrt{1-v^{2}},
\label{(18)}
\end{equation}
where $p>0$, and then introduce a boost in the $Z_{1}$-direction by
defining yet new coordinates $Y_{\mu}$, independent of $v$,
such that (hereafter $\gamma \equiv (1-v^{2})^{-1/2}$)
\begin{equation}
Z_{0}=\gamma (Y_{0}+v Y_{1}),
\label{(19)}
\end{equation}
\begin{equation}
Z_{1}=\gamma (vY_{0}+Y_{1}),
\label{(20)}
\end{equation}
\begin{equation}
Z_{2}=Y_{2}, \; Z_{3}=Y_{3}, \; Z_{4}=Y_{4}.
\label{(21)}
\end{equation}
The metric (13) is therefore re-expressed, eventually, in the form
(see Appendix A)
\begin{eqnarray}
ds^{2}&=& \gamma^{2} (h_{00}+v^{2})
dY_{0}^{2}+\gamma^{2}(1+v^{2}h_{00})dY_{1}^{2} 
+dY_{2}^{2}+dY_{3}^{2}+h_{44}dY_{4}^{2} \nonumber \\
&+& 2v \gamma^{2}(1+h_{00})dY_{0}dY_{1}+2\gamma h_{04}dY_{0}dY_{4} 
+ 2v \gamma h_{04}dY_{1}dY_{4},
\label{(22)}
\end{eqnarray}
with the understanding that Eqs. (10)--(12) and (19)--(21) should be inserted
into Eqs. (14)--(17) to express the metric components in Eq. (22) 
completely in $Y_{\mu}$ coordinates (i.e. $h_{ab}=h_{ab}(Y_{\mu})$ 
in Eq. (22)).

\section{Schwarzschild-anti de Sitter space-time}

To complete our starting set of formulae 
we now consider Schwarzschild-anti de Sitter
space-time \cite{Podo97}, whose metric takes originally the form (cf. 
Eq. (2))
\begin{equation}
ds^{2}=-\left(1-{2m \over r}+{r^{2}\over a^{2}}\right)dt^{2}
+{dr^{2}\over \left(1-{2m \over r}+{r^{2}\over a^{2}}\right)}
+r^{2}(d\theta^{2}+\sin^{2}\theta d\phi^{2}).
\label{(23)}
\end{equation}
Such a geometry can be represented as the four-dimensional hyperboloid
\begin{equation}
-Z_{0}^{2}+Z_{1}^{2}+Z_{2}^{2}+Z_{3}^{2}-Z_{4}^{2}=-a^{2},
\label{(24)}
\end{equation}
embedded in a five-dimensional space-time with metric
\begin{equation}
ds^{2}=-dZ_{0}^{2}+dZ_{1}^{2}+dZ_{2}^{2}+dZ_{3}^{2}-dZ_{4}^{2},
\label{(25)}
\end{equation}
corresponding to two timelike directions, i.e. $Z_{0}$ and $Z_{4}$.
The natural parametrization of this is given by \cite{Podo97}
\begin{equation}
Z_{0} \equiv \sqrt{a^{2}+r^{2}} \; \sin(t/a),
\label{(26)}
\end{equation}
\begin{equation}
Z_{4} \equiv \sqrt{a^{2}+r^{2}} \; \cos(t/a),
\label{(27)}
\end{equation}
while $Z_{1},Z_{2}$ and $Z_{3}$ remain defined as in Eqs. (7)--(9).
Thus, on further defining (cf. Eqs. (10) and (11))
\begin{equation}
{\widetilde f}^{2} \equiv a^{2}+r^{2} = Z_{0}^{2}+Z_{4}^{2},
\label{(28)}
\end{equation}
\begin{equation}
{\widetilde F}_{m} \equiv 1-{2a^{2}m \over {\widetilde f}^{2}r}
+{a^{2} / r^{2} \over \left(1-{2a^{2}m \over {\widetilde f}^{2}r}
\right)},
\label{(29)}
\end{equation}
we first re-express the metric (23) in the form
\begin{equation}
ds^{2}=H_{00}dZ_{0}^{2}+H_{44}dZ_{4}^{2}+2H_{04}dZ_{0}dZ_{4}
+dZ_{1}^{2}+dZ_{2}^{2}+dZ_{3}^{2},
\label{(30)}
\end{equation}
where
\begin{equation}
H_{00} \equiv {Z_{0}^{2} \over {\widetilde f}^{2}}{\widetilde F}_{m}
-\left(1-{2a^{2}m \over {\widetilde f}^{2}r}\right)
-{Z_{0}^{2}\over r^{2}},
\label{(31)}
\end{equation}
\begin{equation}
H_{44} \equiv \left(1-{Z_{0}^{2}\over {\widetilde f}^{2}}\right)
{\widetilde F}_{m}-\left(1-{2a^{2}m \over {\widetilde f}^{2}r}\right)
-{Z_{4}^{2}\over r^{2}},
\label{(32)}
\end{equation}
\begin{equation}
H_{04} \equiv {Z_{0}Z_{4} \over {\widetilde f}^{2}}
{\widetilde F}_{m}-{Z_{0}Z_{4}\over r^{2}}.
\label{(33)}
\end{equation}
Now the boost in the $Z_{1}$-direction is again given by Eqs. (18)--(21)
\cite{Podo97}, so that, in full analogy with Eq. (22), 
the boosted metric reads
\begin{equation}
{\widetilde H}_{00}=\gamma^{2}(H_{00}+v^{2}),
\label{(34)}
\end{equation}
\begin{equation}
{\widetilde H}_{11}=\gamma^{2}(1+v^{2}H_{00})
=1+v^{2}+v^{2}{\widetilde H}_{00},
\label{(35)}
\end{equation}
\begin{equation}
{\widetilde H}_{22}={\widetilde H}_{33}=1,
\label{(36)}
\end{equation}
\begin{equation}
{\widetilde H}_{44}=H_{44},
\label{(37)}
\end{equation}
\begin{equation}
{\widetilde H}_{01}=v \gamma^{2} (1+H_{00})=v(1+{\widetilde H}_{00}),
\label{(38)}
\end{equation}
\begin{equation}
{\widetilde H}_{04}=\gamma H_{04},
\label{(39)}
\end{equation}
\begin{equation}
{\widetilde H}_{14}=v \gamma H_{04}.
\label{(40)}
\end{equation}
In these formulae, the `boosted' $H_{00},H_{44}$ and $H_{04}$ components
can be written, upon defining (the subscript `minus' refers here to the
negative cosmological constant)
\begin{equation}
W_{-}(a,v,Y) \equiv 
{1\over (Y_{0}+vY_{1})^{2}+(1-v^{2})(Y_{4}^{2}-a^{2})},
\label{(41)}
\end{equation}
\begin{eqnarray}
\rho_{-}(a,v,Y)& \equiv & W_{-}(0,v,Y)-W_{-}(a,v,Y)
-2a^{2}p(1-v^{2})^{2}W_{-}^{2}(0,v,Y)W_{-}^{1/2}(a,v,Y) \nonumber \\
&+& {a^{2}(1-v^{2})W_{-}(0,v,Y)W_{-}(a,v,Y) \over
1-2a^{2}p (1-v^{2})^{2}W_{-}(0,v,Y)W_{-}^{1/2}(a,v,Y)},
\label{(42)}
\end{eqnarray}
in the form 
\begin{equation}
H_{00}=(Y_{0}+vY_{1})^{2}\rho_{-}(a,v,Y)
-\Bigr(1-2a^{2}p (1-v^{2})^{2}W_{-}(0,v,Y)W_{-}^{1/2}(a,v,Y)\Bigr),
\label{(43)}
\end{equation}
\begin{equation}
H_{44}=(1-v^{2})Y_{4}^{2}\rho_{-}(a,v,Y)
-\Bigr(1-2a^{2}p(1-v^{2})^{2}W_{-}(0,v,Y)W_{-}^{1/2}(a,v,Y)\Bigr),
\label{(44)}
\end{equation}
\begin{equation}
H_{04}={1\over \gamma}(Y_{0}+vY_{1})Y_{4}\rho_{-}(a,v,Y),
\label{(45)}
\end{equation}
where we have exploited Eqs. (28), (29) and (31)--(33). The advantage of
Eqs. (34)--(45) is that they are non-perturbative, exact formulae for
the boosted metric that relate the analysis of the 
singularity structure to one function
only, i.e. $W_{-}(a,v,Y)$ defined in Eq. (41).

\section{Shock-wave geometry from the ultrarelativistic limit}

In the Schwarzschild-de Sitter geometry of Secs. II and III 
we find in analogous way,
on defining (the subscript `plus' refers here to the positive
cosmological constant)
\begin{equation}
W_{+}(a,v,Y) \equiv {1\over (Y_{0}+vY_{1})^{2}
+(1-v^{2})(a^{2}-Y_{4}^{2})},
\label{(46)}
\end{equation}
\begin{eqnarray}
\rho_{+}(a,v,Y)& \equiv & W_{+}(0,v,Y)-W_{+}(a,v,Y)
+2a^{2}p(1-v^{2})^{2}W_{+}^{2}(0,v,Y)W_{+}^{1/2}(a,v,Y) \nonumber \\
&-& {a^{2}(1-v^{2})W_{+}(0,v,Y)W_{+}(a,v,Y) \over
1+2a^{2}p(1-v^{2})^{2}W_{+}(0,v,Y)W_{+}^{1/2}(a,v,Y)},
\label{(47)}
\end{eqnarray}
the basic formulae
\begin{equation}
h_{00}=(Y_{0}+vY_{1})^{2}\rho_{+}(a,v,Y)
-\Bigr(1+2a^{2}p(1-v^{2})^{2}W_{+}(0,v,Y)W_{+}^{1/2}(a,v,Y)\Bigr),
\label{(48)}
\end{equation}
\begin{equation}
h_{44}=(1-v^{2})Y_{4}^{2}\rho_{+}(a,v,Y)
+\Bigr(1+2a^{2}p(1-v^{2})^{2}W_{+}(0,v,Y)W_{+}^{1/2}(a,v,Y)\Bigr),
\label{(49)}
\end{equation}
\begin{equation}
h_{04}=-{1\over \gamma}(Y_{0}+vY_{1}) Y_{4} \rho_{+}(a,v,Y).
\label{(50)}
\end{equation}

At this stage, we exploit the fundamental identity \cite{Hott93}
(here $f$ is any summable function on the real line)
\begin{equation}
\lim_{v \to 1} \gamma f \Bigr(\gamma^{2}(Y_{0}+vY_{1})^{2} \Bigr)
=\delta (Y_{0}+Y_{1}) \int_{-\infty}^{\infty}f(x^{2})dx,
\label{(51)}
\end{equation}
and add carefully the various terms in Eq. (47) to find that,
on defining the even functions of $x$ (hereafter $a> Y_{4}$)
\begin{equation}
f_{1}(x) \equiv (x^{2}-Y_{4}^{2})\sqrt{x^{2}+a^{2}-Y_{4}^{2}}
\Bigr[(x^{2}-Y_{4}^{2})(x^{2}+a^{2}-Y_{4}^{2})
+2a^{2}p\sqrt{1-v^{2}}\sqrt{x^{2}+a^{2}-Y_{4}^{2}}\Bigr],
\label{(52)}
\end{equation}
\begin{equation}
f_{2}(x) \equiv (x^{2}-Y_{4}^{2})^{2} \sqrt{x^{2}+a^{2}-Y_{4}^{2}},
\label{(53)}
\end{equation}
\begin{equation}
f_{3}(x) \equiv (x^{2}-Y_{4}^{2})\sqrt{x^{2}+a^{2}-Y_{4}^{2}},
\label{(54)}
\end{equation}
one has (hereafter $x_{v} \equiv \gamma (Y_{0}+vY_{1})$)
\begin{equation}
\rho_{+}(a,v,Y)=2a^{2}p \gamma \left({a^{2}\over f_{1}(x_{v})}
+{1\over f_{2}(x_{v})}\right),
\label{(55)}
\end{equation}
and hence, by virtue of the limit (51) (with our notation, 
${\widetilde h}_{ab}$ is obtained from $h_{ab}$ exactly as 
${\widetilde H}_{ab}$ is obtained from $H_{ab}$ in Eqs. (34)--(40),
as is clear from Eq. (22)), 
\begin{equation}
\lim_{v \to 1}{\widetilde h}_{00}=-1+2a^{2}p \delta(Y_{0}+Y_{1})
\lim_{v \to 1} \int_{-\infty}^{\infty}
\left[{a^{2}x^{2}\over f_{1}(x)}+{x^{2}\over f_{2}(x)}
-{1\over f_{3}(x)}\right]dx,
\label{(56)}
\end{equation}
\begin{equation}
\lim_{v \to 1}{\widetilde h}_{44}=1+2a^{2}p \delta(Y_{0}+Y_{1})
\lim_{v \to 1}(1-v^{2}) \int_{-\infty}^{\infty}
\left[Y_{4}^{2}\left({a^{2}\over f_{1}(x)}+{1\over f_{2}(x)}\right)
+{1\over f_{3}(x)}\right]dx =1,
\label{(57)}
\end{equation}
\begin{equation}
\lim_{v \to 1}{\widetilde h}_{04}=-2a^{2}pY_{4} \lim_{v \to 1}
\sqrt{1-v^{2}} \int_{-\infty}^{\infty} x
\left({a^{2}\over f_{1}(x)}+{1\over f_{2}(x)}\right)dx=0.
\label{(58)}
\end{equation}
In Eq. (56), the desired limit can be brought within the integral, and
the $v$-dependent part of $f_{1}(x)$ gives vanishing contribution to
this limit. One thus finds
\begin{eqnarray}
\; & \; & \lambda_{+} \equiv \lim_{v \to 1} \int_{-\infty}^{\infty}
\left[{a^{2}x^{2}\over f_{1}(x)}+{x^{2}\over f_{2}(x)}
-{1\over f_{3}(x)}\right]dx \nonumber \\
&=& 2 \left[(a^{2}+Y_{4}^{2})+(a^{2}-Y_{4}^{2})
a{\partial \over \partial a} \right]
\int_{0}^{\infty}
{dx\over (x^{2}-Y_{4}^{2})^{2}(x^{2}+a^{2}-Y_{4}^{2})^{1/2}}.
\label{(59)}
\end{eqnarray}
In this integral we now change integration variable so as to get rid of
square roots in the integrand, by defining
\begin{equation}
\tau \equiv x+\sqrt{x^{2}+a^{2}-Y_{4}^{2}},
\label{(60)}
\end{equation}
which implies that
\begin{equation}
{dx \over \sqrt{x^{2}+a^{2}-Y_{4}^{2}}}={d \tau \over \tau},
\label{(61)}
\end{equation}
\begin{equation}
x^{2}-Y_{4}^{2}={g(\tau)\over 4\tau^{2}},
\label{(62)}
\end{equation}
where (hereafter both $(a-Y_{4})$ and $(a+Y_{4})$ are 
taken to be positive)
\begin{equation}
g(\tau) \equiv (\tau^{2}-(a^{2}-Y_{4}^{2}))^{2}-(2\tau Y_{4})^{2}
=(\tau-(Y_{4}+a))(\tau-(Y_{4}-a))(\tau+Y_{4}+a)(\tau-(a-Y_{4})).
\label{(63)}
\end{equation}
Hence we find
\begin{eqnarray}
\lambda_{+}&=& 32 \left[(a^{2}+Y_{4}^{2})+(a^{2}-Y_{4}^{2})
a {\partial \over \partial a} \right]
\int_{\sqrt{a^{2}-Y_{4}^{2}}}^{\infty}
{\tau^{3}\over g^{2}(\tau)}d\tau \nonumber \\
&=& \left[(a^{2}+Y_{4}^{2})+(a^{2}-Y_{4}^{2})
a {\partial \over \partial a}\right]
\left[-{1\over (aY_{4})^{2}}+{(a^{2}+Y_{4}^{2})\over 2 (aY_{4})^{3}}
\log \left({{a+Y_{4}}\over {a-Y_{4}}}\right)\right] \nonumber \\
&=& -{4\over a^{2}}+{2Y_{4}\over a^{3}}
\log \left({{a+Y_{4}}\over {a-Y_{4}}}\right).
\label{(64)}
\end{eqnarray}

Furthermore, in Eq. (57), 
the desired limit vanishes, since the integral therein is
finite and is multiplied by a vanishing function of $v$ 
(i.e. $(1-v^{2})$) as $v \rightarrow 1$, while the limit in Eq. (58)
vanishes because $\sqrt{1-v^{2}}$ vanishes as well
as the integral therein (being the integral of an odd function over 
the whole real line).
In agreement with Ref. \cite{Hott93}, we therefore obtain,
from Eqs. (22), (56) and (64), the singular
boosted metric (cf. Eq. (A17))
\begin{eqnarray}
ds^{2}&=& -dY_{0}^{2}+\sum_{i=1}^{4}dY_{i}^{2} \nonumber \\
&+& 4p \left[-2+{Y_{4}\over a}\log \left({{a+Y_{4}}\over {a-Y_{4}}}\right) 
\right] \delta(Y_{0}+Y_{1}) \; (dY_{0}+dY_{1})^{2},
\label{(65)}
\end{eqnarray}
i.e. de Sitter space plus a shock-wave singularity located on the
null hypersurface described by the equations
\begin{equation}
Y_{0}+Y_{1}=0, \;
Y_{2}^{2}+Y_{3}^{2}+(Y_{4}^{2}-a^{2})=0.
\label{(66)}
\end{equation}

In an entirely analogous way, the ultrarelativistic limit of
Schwarzschild-anti de Sitter space-time can be obtained,
after adding carefully the terms in Eq. (42), by defining the
even functions of $x$ (hereafter $Y_{4}>a$)
\begin{equation}
\varphi_{1}(x) \equiv (x^{2}+Y_{4}^{2})\sqrt{x^{2}+Y_{4}^{2}-a^{2}}
\Bigr[(x^{2}+Y_{4}^{2})(x^{2}+Y_{4}^{2}-a^{2})
-2a^{2}p\sqrt{1-v^{2}}\sqrt{x^{2}+Y_{4}^{2}-a^{2}}\Bigr],
\label{(67)}
\end{equation}
\begin{equation}
\varphi_{2}(x) \equiv (x^{2}+Y_{4}^{2})^{2} \sqrt{x^{2}+Y_{4}^{2}-a^{2}},
\label{(68)}
\end{equation}
\begin{equation}
\varphi_{3}(x) \equiv (x^{2}+Y_{4}^{2})\sqrt{x^{2}+Y_{4}^{2}-a^{2}},
\label{(69)}
\end{equation}
which occur in the identity
\begin{equation}
\rho_{-}(a,v,Y)=2a^{2}p \gamma \left({a^{2}\over \varphi_{1}(x_{v})}
-{1\over \varphi_{2}(x_{v})}\right).
\label{(70)}
\end{equation}
By exploiting again the limit (51) we therefore obtain
\begin{equation}
\lim_{v \to 1}{\widetilde H}_{00}=-1+2a^{2}p \delta(Y_{0}+Y_{1})
\lim_{v \to 1} \int_{-\infty}^{\infty}
\left[{a^{2}x^{2}\over \varphi_{1}(x)}-{x^{2}\over \varphi_{2}(x)}
+{1\over \varphi_{3}(x)}\right]dx,
\label{(71)}
\end{equation}
\begin{equation}
\lim_{v \to 1}{\widetilde H}_{44}=-1+2a^{2}p \delta(Y_{0}+Y_{1})
\lim_{v \to 1}(1-v^{2}) \int_{-\infty}^{\infty}
\left[Y_{4}^{2}\left({a^{2}\over \varphi_{1}(x)}
-{1\over \varphi_{2}(x)}\right)
+{1\over \varphi_{3}(x)}\right]dx =-1,
\label{(72)}
\end{equation}
\begin{equation}
\lim_{v \to 1}{\widetilde H}_{04}=2a^{2}pY_{4} \lim_{v \to 1}
\sqrt{1-v^{2}} \int_{-\infty}^{\infty} x
\left({a^{2}\over \varphi_{1}(x)}-{1\over \varphi_{2}(x)}\right)dx=0.
\label{(73)}
\end{equation}
In Eq. (71), the desired limit can be brought within the integral as in
Eq. (56), and the $v$-dependent part of $\varphi_{1}(x)$ gives vanishing
contribution to this limit. One thus finds 
\begin{eqnarray}
\; & \; & \lambda_{-} \equiv \lim_{v \to 1}\int_{-\infty}^{\infty}
\left[{a^{2}x^{2}\over \varphi_{1}(x)}-{x^{2}\over \varphi_{2}(x)}
+{1\over \varphi_{3}(x)}\right]dx \nonumber \\
&=& 2 \left[(a^{2}+Y_{4}^{2})+(a^{2}-Y_{4}^{2})
a{\partial \over \partial a} \right]
\int_{0}^{\infty}{dx \over (x^{2}+Y_{4}^{2})^{2}
(x^{2}+Y_{4}^{2}-a^{2})^{1\over 2}}.
\label{(74)}
\end{eqnarray}
In this integral, we now change integration variables according to
(cf. Eq. (60))
\begin{equation}
T \equiv x+\sqrt{x^{2}+Y_{4}^{2}-a^{2}},
\label{(75)}
\end{equation}
which implies that
\begin{equation}
{dx \over \sqrt{x^{2}+Y_{4}^{2}-a^{2}}}={dT \over T},
\label{(76)}
\end{equation}
\begin{equation}
x^{2}+Y_{4}^{2}={G(T) \over 4T^{2}},
\label{(77)}
\end{equation}
where (hereafter both $(Y_{4}-a)$ and $(Y_{4}+a)$ are 
taken to be positive)
\begin{equation}
G(T) \equiv (T^{2}-(Y_{4}^{2}-a^{2}))^{2}-(2iTY_{4})^{2}
=(T-i(Y_{4}+a))(T-i(Y_{4}-a))(T+i(Y_{4}+a))(T-i(a-Y_{4})).
\label{(78)}
\end{equation}
Hence we find
\begin{eqnarray}
\lambda_{-} &=& 32 \left[(a^{2}+Y_{4}^{2})+(a^{2}-Y_{4}^{2})
a {\partial \over \partial a}\right]
\int_{\sqrt{Y_{4}^{2}-a^{2}}}^{\infty}{T^{3}\over G^{2}(T)}dT
\nonumber \\
&=& -{4\over a^{2}}+{2Y_{4}\over a^{3}}
\log \left({{Y_{4}+a}\over {Y_{4}-a}}\right).
\label{(79)}
\end{eqnarray}
The resulting singular boosted metric is, from Eqs. (34)-(40),
(71) and (79),
\begin{eqnarray}
ds^{2}&=& -dY_{0}^{2}+dY_{1}^{2}+dY_{2}^{2}+dY_{3}^{2}-dY_{4}^{2}
\nonumber \\
&+& 4p \left[-2+{Y_{4}\over a}\log \left(
{{Y_{4}+a}\over {Y_{4}-a}}\right)\right] 
\delta(Y_{0}+Y_{1}) 
\; (dY_{0}+dY_{1})^{2},
\label{(80)}
\end{eqnarray}
i.e. anti-de Sitter space plus a shock-wave singularity located on the
null hypersurface described by the equations (cf. Eq. (66))
\begin{equation}
Y_{0}+Y_{1}=0, \; Y_{2}^{2}+Y_{3}^{2}-(Y_{4}^{2}-a^{2})=0.
\label{(81)}
\end{equation}

By inspection of Eqs. (64) and (79) we notice that $\lambda_{-}$ is not
obtained from $\lambda_{+}$ by the replacements $a \rightarrow \pm ia$,
$Y_{4} \rightarrow \pm iY_{4}$, since the terms $-{4\over a^{2}}$ and
${2Y_{4}\over a^{3}}$ remain the same. 
Note also that, at $Y_{4}=a$, the limits (59) and (74) reduce to
\begin{equation}
\lambda_{\pm}=2a^{2}\int_{-\infty}^{\infty}
{dx \over x (x^{2} \mp a^{2})^{2}},
\label{(82)}
\end{equation}
which vanishes, being the integral of an odd function over the whole
real line. Thus, only $|a-Y_{4}| >0$ concerns the shock-wave geometry.

\section{Concluding remarks}

Although our final result agrees with the findings in Ref. \cite{Hott93},
our work is original and of interest for at least two reasons:
\vskip 0.3cm
\noindent
(i) We have performed an exact, fully non-perturbative analysis of boosted
space-times with cosmological constant, dealing at all stages with the 
whole set of non-linearities of the metric. This offers some advantages
from the point of view of both physics and mathematics: 
for any fixed value of $v \not =1$, the only reliable 
formulae for the metric are our Eqs. (13)--(17), (22), (30)--(50).
Moreover, although the linearized metric is sufficient to derive the 
ultrarelativistic limit, it is instructive to understand how the final
results come to agree in a background with cosmological constant.  
\vskip 0.3cm
\noindent
(ii) We have provided detailed analytic formulae also for the 
Schwarzschild-anti-de Sitter space-time.

Our investigation is therefore part of the efforts
aimed at a better understanding
of the circumstances under which one can predict formation of
shock-wave geometries \cite{Stew06} in classical or quantum gravity
\cite{Cai98, Cai99}.  
In future work, we hope to be able to study the `boosted' Riemann tensor,
along the lines of Ref. \cite{Aich71}, but for Schwarzschild-de Sitter,
Schwarzschild-anti de Sitter (see previous sections) 
and Kerr--Schild geometries \cite{Ferr90}. Moreover, 
applications (if any) of our exact analysis to quantum gravity effects
in space-times with cosmological constant should be investigated.

\acknowledgments 
The work of G. Esposito was partially supported by PRIN {\it SINTESI}.
R. Pettorino gratefully acknowledges G. Veneziano for the hospitality
at College de France where part of this work was done. The work of
Roberto Pettorino has been partially supported by the European 
Community's Human Potential Programme under contract 
MRTN-CT-2004-005104 ``Constituents, Fundamental Forces and Symmetries
of The Universe''. 

\appendix
\section{The boosted metric}

In the course of arriving at Eq. (13) from Eqs. (2) and (5)--(9), the
relevant intermediate steps are the formulae (here $\varepsilon \equiv
\pm 1, {\rm ch}_{t} \equiv \cosh (t/a), {\rm sh}_{t} \equiv
\sinh (t/a)$)
\begin{equation}
dZ_{0}=-{r\over f}{\rm sh}_{t}dr+{f\over a}{\rm ch}_{t}dt,
\label{(A1)}
\end{equation}
\begin{equation}
dZ_{4}=-{\varepsilon r \over f}{\rm ch}_{t}dr
+{\varepsilon f \over a}{\rm sh}_{t}dt,
\label{(A2)}
\end{equation}
\begin{equation}
dZ_{0}^{2}-dZ_{4}^{2}={f^{2}\over a^{2}}dt^{2}
-{r^{2}\over f^{2}}dr^{2},
\label{(A3)}
\end{equation}
\begin{equation}
dZ_{0}^{2}+dZ_{4}^{2}={r^{2} \over f^{2}}\Bigr(1+2{\rm sh}_{t}^{2}\Bigr)
dr^{2}+{f^{2}\over a^{2}}\Bigr(1+2 {\rm sh}_{t}^{2}\Bigr)dt^{2}
-{4r \over a}{\rm sh}_{t} {\rm ch}_{t}drdt,
\label{(A4)}
\end{equation}
\begin{equation}
dZ_{0}dZ_{4}={\varepsilon r^{2}\over f^{2}}{\rm sh}_{t} {\rm ch}_{t}dr^{2}
+{\varepsilon f^{2}\over a^{2}}{\rm sh}_{t}{\rm ch}_{t}dt^{2}
-{\varepsilon r \over a}\Bigr(1+2{\rm sh}_{t}^{2}\Bigr)dr dt.
\label{(A5)}
\end{equation}
Thus, on defining (see Eqs. (5), (6), (10), (12))
\begin{equation}
p_{t} \equiv {\rm ch}_{t} {\rm sh}_{t},
\label{(A6)}
\end{equation}
\begin{equation}
q_{t} \equiv 1+2{\rm sh}_{t}^{2}
-{4 p_{t}^{2} \over (1+2 {\rm sh}_{t}^{2})}
=(1+2{\rm sh}_{t}^{2})^{-1}=Q^{-1},
\label{(A7)}
\end{equation}
one arrives at
\begin{equation}
dt^{2}= {a^{2}\over f^{2}}\left[{1\over 2}\left({1 \over q_{t}}+1 \right)
dZ_{0}^{2}+{1\over 2}\left({1 \over q_{t}}-1 \right)dZ_{4}^{2}
- {2\over \varepsilon} p_{t} dZ_{0}dZ_{4}\right],
\label{(A8)}
\end{equation}
\begin{equation}
dr^{2}={f^{2}\over r^{2}}\left[{1\over 2}\left({1 \over q_{t}}-1 \right)
dZ_{0}^{2}+{1\over 2}\left({1 \over q_{t}}+1 \right)dZ_{4}^{2}
-{2\over \varepsilon} p_{t} dZ_{0}dZ_{4}\right],
\label{(A9)}
\end{equation}
from which a little amount of calculations yields Eqs. (13)--(17).
In the course of deriving Eq. (22), we have re-expressed Eq. (A9) 
in the form
\begin{equation}
dr^{2}={Z_{0}^{2}\over r^{2}}dZ_{0}^{2}+{Z_{4}^{2}\over r^{2}}
dZ_{4}^{2}-{2Z_{0}Z_{4}\over r^{2}}dZ_{0}dZ_{4},
\label{(A10)}
\end{equation}
while bearing in mind that
\begin{equation}
r^{2}(d\theta^{2}+\sin^{2}\theta d\phi^{2})=dZ_{1}^{2}+dZ_{2}^{2}+dZ_{3}^{2}
-dr^{2}.
\label{(A11)}
\end{equation}

Along similar lines, we exploit in Sec. IV the identities
\begin{equation}
dt^{2}={a^{2}\over {\widetilde f}^{2}}\left[
{Z_{4}^{2}\over {\widetilde f}^{2}}dZ_{0}^{2}
+{Z_{0}^{2}\over {\widetilde f}^{2}}dZ_{4}^{2}
-{2Z_{0}Z_{4}\over {\widetilde f}^{2}}\right],
\label{(A12)}
\end{equation}
\begin{equation}
dr^{2}={Z_{0}^{2}\over r^{2}}dZ_{0}^{2}+{Z_{4}^{2}\over r^{2}}
dZ_{4}^{2}+{2Z_{0}Z_{4}\over r^{2}}dZ_{0}dZ_{4}.
\label{(A13)}
\end{equation}

The procedure for deriving a shock wave metric from a solution of the
Einstein equations considers initially the metric for the latter
in the form \cite{Deve89}
\begin{equation}
ds^{2}=2A(u,v')dudv'+g_{ij}dx^{i}dx^{j}.
\label{(A14)}
\end{equation}
Let now $w$ be a new coordinate defined by ($\Theta$ being the step
function)
\begin{equation}
w \equiv v'+f(X^{i})\Theta(u),
\label{(A15)}
\end{equation}
so that
\begin{equation}
dv'=dw-f(X^{i})\delta(u)du,
\label{(A16)}
\end{equation}
where we have exploited the distributional relation $\Theta'(u)=\delta(u)$.
The metric (A14) takes therefore the shock-wave form
\begin{equation}
ds^{2}=2A(u,w)dudw-2A(u,w)f(X^{i})\delta(u)du^{2}
+g_{ij}dx^{i}dx^{j}.
\label{(A17)}
\end{equation}

\end{document}